\documentclass[letterpaper, 10 pt, conference]{ieeeconf}  % Comment this line out if you need a4paper

\IEEEoverridecommandlockouts                        
\usepackage{kotex}
\usepackage{graphicx}
\usepackage{color,soul}
\usepackage{bm,upgreek}
\usepackage{algorithm}
\usepackage{amsmath,amssymb,amsfonts}
\usepackage{algpseudocode}
\usepackage{multirow}
\usepackage{breqn}
\usepackage{gensymb}
\usepackage{hyperref}
\usepackage{cite}
\usepackage{caption}
\usepackage{subcaption}
\renewcommand{\QED}{\QEDopen}
\newtheorem{definition}{Definition}
\newtheorem{assumption}{Assumption}
\newtheorem{theorem}{Theorem}
\newtheorem{remark}{Remark}
\usepackage{cleveref}
\crefname{assumption}{assumption}{assumptions}
\usepackage{diagbox}
\usepackage{siunitx}

\title{\LARGE \bf Parameter-Varying Koopman Operator for Nonlinear System Modeling and Control}
\author{Changyu Lee, Kiyong Park, and Jinwhan Kim % <-this % stops a space
\thanks{This research was supported by a grant from National R\&D Project "Development of an electric-powered car ferry and a roll-on/roll-off power supply system" funded by Ministry of Oceans and Fisheries, Korea (PMS5530). \textit{(Corresponding author: Jinwhan Kim)}}% <-this % stops a space
\thanks{The authors are with the Department of Mechanical Engineering, Korea Advanced Institute of Science and Technology, Daejeon 34141, South Korea (e-mail: leeck@kaist.ac.kr; qkrrldyd777@kaist.ac.kr; jinwhan@kaist.ac.kr)}%
}

\begin{document}
\maketitle
\thispagestyle{empty}
\pagestyle{empty}

\everymath{\displaystyle}
\begin{abstract}
This paper proposes a novel approach for modeling and controlling nonlinear systems with varying parameters. The approach introduces the use of a parameter-varying Koopman operator (PVKO) in a lifted space, which provides an efficient way to understand system behavior and design control algorithms that account for underlying dynamics and changing parameters. The PVKO builds on a conventional Koopman model by incorporating local time-invariant linear systems through interpolation within the lifted space. This paper outlines a procedure for identifying the PVKO and designing a model predictive control using the identified PVKO model. Simulation results demonstrate that the proposed approach improves model accuracy and enables predictions based on future parameter information. The feasibility and stability of the proposed control approach are analyzed, and their effectiveness is demonstrated through simulation.
\end{abstract}


\section{Introduction}
Model predictive control (MPC) is a powerful algorithm that has proven to be effective for controlling nonlinear systems in various applications, including robotics and transportation \cite{mayne2014model,kim2023navigable,lee2023nonlinear}. MPC offers several advantages, such as the ability to handle state and input constraints and the capacity to tackle multi-input multi-output nonlinear systems. However, nonlinear systems pose challenges in optimizing control due to their non-convex nature, resulting in computational complexity and difficulties in ensuring stability and robustness. Additionally, unreliable models can lead to performance degradation and system failure due to constraint violations \cite{langson2004robust}. Therefore, obtaining accurate system models and addressing non-convex problems are essential for effective MPC, but these tasks can be challenging in practical applications.

Recently, data-driven Koopman operator (KO)-based system identification has gained popularity in research. The KO provides a linear representation of nonlinear autonomous systems in infinite dimensions \cite{koopman1931hamiltonian}, which can further be approximated in a finite number of dimensions through data-driven approaches \cite{brunton2016koopman}. 
In this approach, user-defined lifting functions and extended dynamic mode decomposition (EDMD) methods are often utilized \cite{dmdc,edmd}. Deep neural networks also offer the capability to simultaneously identify lifting functions as well as the KO \cite{deepkoopman}.
By incorporating a linear MPC algorithm into the linear system in the lifted dimension, the approach can be executed to nonlinear MPC \cite{korda2018linear}. Futhermore, robust MPC has been developed to address model uncertainty resulting from the identification process based on KOs \cite{zhang2022robust, mamakoukas2022robust}. These findings suggest the potential of the KO-based approach to address non-convex problems. 
However, the success of data-driven identification methods heavily depends on the quantity and quality of data, and challenges still remain in this area.
 
\begin{figure}[t]
    \centering
    \includegraphics[width=0.95\linewidth]{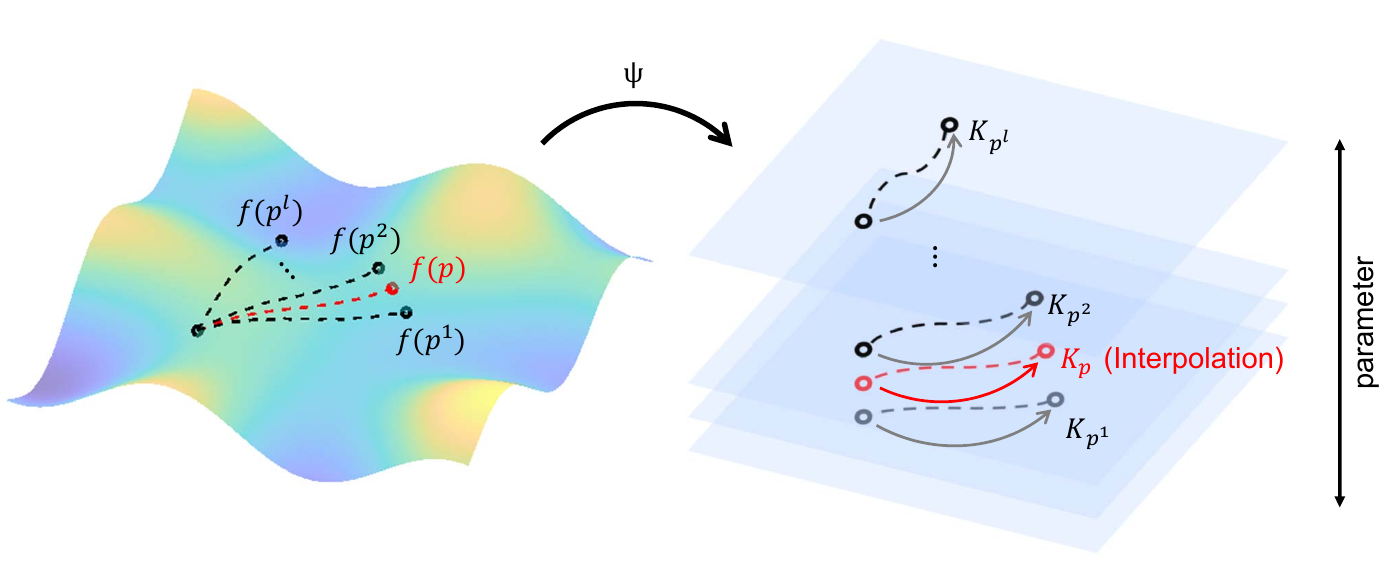}
    \caption{An illustration of the proposed parameter-varying Koopman operator.}
    \label{fig:my_label}
\end{figure}

In previous research, linear time-invariant models have often been used to represent nonlinear systems in the lifted space. However, in many real-world systems, the dynamics are dependent on the operating point. For instance, the lateral dynamics of vehicles are influenced by speed, and chemical process models are highly affected by temperature \cite{LPV_ship,LPV_workingpoint}. To address this issue, linear parameter varying (LPV) or quasi-LPV models have been proposed for modeling and designing control systems \cite{LPV_systems,qLPV_systems,lovera2010guest,hoffmann2014survey,morato2020model}. These models account for the influence of exogenous parameters on the system dynamics and provide a more accurate representation of the system behavior. By considering the dependence of the system dynamics on the operating point, LPV models enable the design of controllers that are more robust and effective.

Motivated by recent advances in LPV systems and identification approaches \cite{zhu2008method, de2014interpolated}, this paper proposes a parameter-varying KO (PVKO) for modeling and controlling nonlinear systems with varying parameters in the lifted space. The proposed approach is based on collecting data for each operating point, identifying a KO for each point, and local interpolation between the KOs is conducted following the approach in \cite{zhu2008method}. 
The resulting PVKO provides an accurate and predictable model that accounts for the underlying dynamics and varying parameters.
To synthesize the control system, the LPV-MPC approach \cite{morato2020model} is used with the PVKO, assuming the predictability of future parameters.
The proposed control system addresses identification uncertainties, and recursive feasibility and stability analysis are provided. Finally, numerical simulations are conducted to verify the effectiveness of the proposed modeling and control approaches.

The rest of this paper is structured as follows. The following section presents the proposed PVKO approach, and Section 3 describes the control system design. The results of the simulations are discussed in Section 4, and the study is concluded in Section 5. 

{\it{Notation}}: The notation $\mathit{I}_{n\times m}$ and $\mathbf{0}_{n\times m}$ denote that $n \times m$ identity and zero matrices, respectively. The Minkowski sum and Pontryagin set difference of two sets $\mathbb{X},\mathbb{Y} \subset \mathbb{R}^n$ are denoted as $\mathbb{X} \oplus \mathbb{Y}$ and $\mathbb{X} \ominus \mathbb{Y}$, respectively. Additionally, Conv$\{\cdot\}$ represents the convex hull formed by the vertices within $\{\cdot\}$.

\section{Parameter-varying Koopman Operator}
Consider the discrete-time nonlinear system defined by:
\begin{equation}
\mathbf{x}_{k+1} = f(\mathbf{x}_{k},\mathbf{u}_{k}),
\end{equation}
where $\mathbf{x}_{k} \in \mathbb{X} \subset \mathbb{R}^n$ and $\mathbf{u}_{k} \in \mathbb{U} \subset \mathbb{R}^m$ denote the state and input vectors, respectively, and the subscript $k$ indicates the time index. Let ${\Psi}(\mathbf{x}_k,\mathbf{u}_k) \in \mathbb{G} : \mathbb{R}^{n+m} \rightarrow \mathbb{R}^{q+m}$ be an observation function that maps the state and input vectors to the lifted space. The observation function can be defined as follows:
\begin{equation}
    {\Psi}(\mathbf{x}_k,\mathbf{u}_k) = \begin{bmatrix}
        \psi_1(\mathbf{x}_{k}),
        \psi_2(\mathbf{x}_{k}),
        \cdots,
        \psi_{q}(\mathbf{x}_{k}),\mathbf{u}_k^\top
    \end{bmatrix}^\top,
\end{equation}
where $\psi_i : \mathbb{R}^n \rightarrow \mathbb{R}$ is the $i$-th component of the observation function.
Then, the lifted state vector can be expressed as follows:
\begin{equation}
\mathbf{y}_{k} = \Psi_{\mathbf{x}}(\mathbf{x}_{k}) = \begin{bmatrix}
        \psi_1(\mathbf{x}_{k}),
        \psi_2(\mathbf{x}_{k}),
        \cdots,
        \psi_{q}(\mathbf{x}_{k})
    \end{bmatrix}^\top,
\end{equation}
where $\mathbf{y}_{k} \in \mathbb{R}^q$ is the lifted state vector.
The KO $\mathcal K: \mathbb{G} \rightarrow \mathbb{G}$ can represent the lifted system in the linear form:
\begin{equation}
  \mathcal{K} (\Psi(\mathbf{x}_{k},\mathbf{u}_{k})) = \Psi (\mathbf{x}_{k+1},\mathbf{u}_{k+1}),
\end{equation}
which can be approximated in a finite-dimensional space higher than $n$ (typically $q \gg n$) using data. Since this approximation is data-driven, a large amount of data is required, and it is necessary to reduce the dimensions $q$ to a manageable level from a control perspective.

In this paper, we focus on a nonlinear system with exogenous parameters defined as follows:
\begin{equation}
\label{eq:nonlinear}
\mathbf{x}_{k+1} = f(\mathbf{x}_{k},\mathbf{u}_{k},p_k),
\end{equation}
where $p_{k} \in \mathbb{P} \subset \mathbb{R}$ is a bounded parameter that introduces uncertainty into the system. To address this, we propose a new approach for modeling the system as a LPV model in a lifted space. Our approach involves using a PVKO $\mathcal{K}_{p_{k}}: \mathbb{G} \rightarrow \mathbb{G}$ defined as:
\begin{align}
  \mathcal{K}_{p_k} (\Psi(\mathbf{x}_{k},\mathbf{u}_{k})) & = \Psi (f(\mathbf{x}_k,\mathbf{u}_k,p_k),\mathbf{u}_{k+1}) \\
  & = \Psi (\mathbf{x}_{k+1},\mathbf{u}_{k+1}),
\end{align}
where $\mathcal{K}_{p_k}$ depends on the parameter. 
 
To identify the PVKO, we use an EDMD-based approach, which involves collecting data from the state and input variables of the system at each working point and using the data to identify the KO for each point. We then use an interpolation-based modeling method to find the PVKO. Let 
\begin{equation}
    \begin{aligned}
        \mathbf{X}(i) &= [\mathbf{x}_1, \mathbf{x}_2, \ldots, \mathbf{x}_{M-1}] \in \mathbb{R}^{n \times (M-1)}, \\
        \mathbf{X}^+(i) &= [\mathbf{x}_2, \mathbf{x}_3, \ldots, \mathbf{x}_{M}] \in \mathbb{R}^{n \times (M-1)}, \\
        \mathbf{U}(i) &= [\mathbf{u}_1, \mathbf{u}_2, \ldots, \mathbf{u}_{M-1}] \in \mathbb{R}^{m \times (M-1)},
    \end{aligned}
\end{equation}
denote the collected state and input data at the $i$-th working point, where $M$ is the number of data points. We then lift the collected data using a lifting function to obtain:
\begin{equation}
    \begin{aligned}
    \mathbf{Y}(i) &= 
    [\mathbf{y}_1, \mathbf{y}_2, \ldots, \mathbf{y}_{M-1}] \in \mathbb{R}^{q \times (M-1)}, \\
    \mathbf{Y}^+(i) &= [\mathbf{y}_2, \mathbf{y}_3, \ldots, \mathbf{y}_{M}] \in \mathbb{R}^{q \times (M-1)}.
    \end{aligned}
\end{equation}
Using the collected data, we can establish the following relations:
\begin{equation}
    \begin{aligned}
    \mathbf{Y}^+(i) &= A(p^i)\mathbf{Y}(i) + B(p^i)\mathbf{U}(i), \\
    \mathbf{X}(i) &=  C\mathbf{Y}(i),
    \end{aligned}
\end{equation}
where $p^i$ represents the parameter at the $i$-th working point and $C$ is the output matrix.
We can then find the state matrix by minimizing the following problems:
\begin{equation}
    \begin{gathered}
        \min_{A(p^i),B(p^i)} ||\mathbf{Y}^+(i)  - (A(p^i)\mathbf{Y}(i) + B(p^i)\mathbf{U}(i))||_F, \\
        \min_{C} ||\mathbf{X}(i)  - C\mathbf{Y}(i)||_F,
    \end{gathered}
\end{equation}
where $||\cdot||_F$ represents the Frobenius norm. 
We can solve these problems analytically using the pseudo-inverse of the matrix $\begin{bmatrix} \mathbf{Y}(i) \ \mathbf{U}(i)\end{bmatrix}^\top$ as follows:
\begin{equation}
    \begin{aligned}
        \begin{bmatrix} A(p^i) & B(p^i)\end{bmatrix}&= \mathbf{Y}^+(i) \begin{bmatrix} \mathbf{Y}(i) \\ \mathbf{U}(i)\end{bmatrix}^\dagger, \\
        C & = \mathbf{X}(i)\mathbf{Y}(i)^\dagger,
    \end{aligned}
    \label{eq:ident_C}
\end{equation}
where $\dagger$ indicates the Moore–Penrose inverse.
To find the pseudo-inverse matrix, we can use the singular value decomposition to decompose $\begin{bmatrix} \mathbf{Y}(i) \ \mathbf{U}(i)\end{bmatrix}^\top$ as follows:
\begin{equation}
    \begin{bmatrix} \mathbf{Y}(i) \\ \mathbf{U}(i)\end{bmatrix} = U \Sigma V^\top. 
\end{equation}
Then, we can approximate $A(p^i)$ and $B(p^i)$ as follows:
\begin{equation}
\begin{aligned}
        \begin{bmatrix} A(p^i) & B(p^i)\end{bmatrix} &\approx \mathbf{Y}^+(i)
        V \Sigma^{-1} U^\top \\
        &= \mathbf{Y}^+(i)
        V \Sigma^{-1} \begin{bmatrix} U_A \ U_B \end{bmatrix},
\end{aligned}
\end{equation}
now we can obtain $A(p^i) \approx \mathbf{Y}^+(i) V \Sigma^{-1} U_A$ and $B(p^i) \approx \mathbf{Y}^+(i) V \Sigma^{-1}U_B$.

For a system with $l\in \mathbb{N}$ working points, we can obtain $l$ different $(A(p^i),B(p^i))$ matrices. The PVKO can then be obtained by interpolating these matrices as follows:
\begin{equation}
\begin{aligned}
    A(p_k) &= \alpha_1(p_k) A(p^1) + \alpha_2(p_k) A(p^2) + \cdots + \alpha_l(p_k) A(p^l), \\
    B(p_k) &= \alpha_1(p_k) B(p^1) + \alpha_2(p_k) B(p^2) + \cdots + \alpha_l(p_k) B(p^l),
\end{aligned}
\end{equation}
where $\alpha_1(p_k), \alpha_2(p_k), \ldots, \alpha_l(p_k)$ are weighting coefficients that depend on the parameter $p_k$. Once we have future parameter information, we can predict the future system matrix using the identified PVKO $(A(p_k),B(p_k))$. This approach allows us to use LPV-MPC \cite{morato2020model}.

\begin{remark} A subsequent identification procedure is required to determine the functional form of the weighting coefficients. In this paper, we use the simplest interpolation technique, linear interpolation, which is cost-effective and can provide adequate results for many applications.
\end{remark}

\section{PVKO-based Model Predictive Control}
\begin{assumption} 
We assumed that the uncertainty of the model approximation, $\mathbf{w}_k$, is unknown and bounded, i.e., $\mathbf{w}_k = \mathbf{y}_{k+1} -(A(p_k)\mathbf{y}_k+B(p_k)\mathbf{u}_k) \in \mathbb{W} \subset \mathbb{R}^q$.
\label{asp:bounded}
\end{assumption} 

We propose a method for synthesizing the LPV-MPC algorithm on the lifted space, named PVKO-MPC, based on the identified PVKO. The LPV system with bounded uncertainty $\mathbf{w}_k$ (as stated in \Cref{asp:bounded}) can be represented in the lifted space as follows:
\begin{equation} 
\begin{aligned}
\label{eq:actual_system}
        \mathbf{y}_{k+1} = A(p_k)\mathbf{y}_k &+ B(p_k)\mathbf{u}_k + \mathbf{w}_k, \\
\text{s.t. }\mathbf{y}_k  &\in \mathbb{Y}, \\
            \mathbf{u}_k  &\in \mathbb{U}, \\
            \mathbf{w}_k  &\in \mathbb{W}.
\end{aligned}
\end{equation}
Let the nominal system of \eqref{eq:actual_system} be represented as:
\begin{equation}
\mathbf{\bar{y}}_{k+1} = A(p_k)\mathbf{\bar{y}}_k + B(p_k)\mathbf{\bar{u}}_k,
\label{eq:nominal_system}
\end{equation}
where $\mathbf{\bar{u}}_k$ and $\mathbf{\bar{y}}_k$ are the nominal input and state vectors that correspond to the system without uncertainty. 
The control input of the system \eqref{eq:actual_system} is then designed as follows:
\begin{equation}
 \mathbf{u}_k= \mathbf{\bar{u}}_k + K(\mathbf{y}_k-\mathbf{\bar{y}}_k),
 \label{eq:TMPC_control}
\end{equation}
where the second term in \eqref{eq:TMPC_control} is the auxiliary state feedback control that compensates for the error.

\begin{definition} [Robust positively invariant set]
A set $\Omega$ is a robust positively invariant (RPI) set of the system $\mathbf{e}_{k+1} = (A(p_k)+B(p_k)K)\mathbf{e}_k + \mathbf{w}_k$, if $(A(p_k)+B(p_k)K)\mathbf{e}_k+\mathbf{w}_k \in \Omega $ for all $\mathbf{e}_k \in \Omega$, $p_k \in \mathbb{P}$, and $\mathbf{w}_k \in \mathbb{W}$.
\end{definition}

\begin{definition} [Quadratic stability]
The system $\mathbf{y}_{k+1} = A^c(p_k)\mathbf{y}_k$ is quadratically stable if there exists $P>0$ such that $ A^c(p_k)^\top P A^c(p_k) - P \leq -Q-K^\top R K$ for all $p_k\in \mathbb{P}$, where $A^c(p_k) = A(p_k)+B(p_k)K$.
\end{definition}

\subsection{Unceratinty compensation and RPI set calculation}
Let the error vector be described by $\mathbf{e}_k = \mathbf{y}_k - \mathbf{\bar{y}}_k$. The error system can be represented using \eqref{eq:actual_system}-\eqref{eq:TMPC_control} as follows:
\begin{equation}
\label{pre:error}
\begin{aligned} 
\mathbf{e}_{k+1} & = A(p_k)(\mathbf{y}_k-\mathbf{\bar{y}}_k) + B(p_k)(\mathbf{u}_k - \mathbf{\bar{u}}_k) + \mathbf{w}_k \\ 
& = (A(p_k) + B(p_k)K) \mathbf{e}_k + \mathbf{w}_k \\
& = A^c(p_k) \mathbf{e}_k + \mathbf{w}_k.
\end{aligned}
\end{equation}

\begin{assumption} \label{ams:qs}
    The system \eqref{pre:error} is quadratically stable.
\end{assumption}

Under the \Cref{ams:qs}, the state feedback controller that minimizes the worst-case cost can be obtained by solving the following semidefinite programming problem:
\begin{equation}
    \begin{gathered} \label{eq:LMI}
        \min_{P, K} \text{tr}(P) \\
        \text {s.t.} \ A^c(p^i)^\top PA^c(p^i)-P \leq -Q-K^\top R K, \\
        \qquad \qquad \qquad \qquad \qquad \qquad \qquad \qquad \text{for} \ i =1,2,\ldots,l,
    \end{gathered}
\end{equation}
where $Q, R$ are weight matrices.
We can transform the optimization problem \eqref{eq:LMI} into the following problem using the Schur complement as follows:
\begin{equation}
\label{eq:lmi_original}
\begin{bmatrix}
P-Q-K^\top RK  & A^c(p^i)^\top \\
A^c(p^i) & P^{-1}
\end{bmatrix} \geq 0, \ \text{for} \ i =1,2,\ldots,l.
\end{equation}
Then, by performing a congruence transformation with $S=P^{-1}$ and introducing $Y=K S$ \cite{boyd1994linear}, we can transform the problem into the following form:
\begin{equation}
\begin{gathered}
\max_{S, Y} \text{tr}(S) \\
\text {s.t.} \\
\begin{bmatrix}
S & SA(p^i)^\top+ Y^\top B^\top & SQ^{1/2} & Y^\top R^{1/2} \\
A(p^i)S+BY & S & \mathbf{0}_{q\times q}  & \mathbf{0}_{q\times m}  \\
Q^{1/2}S & \mathbf{0}_{q\times q}  & \mathit{I}_{q\times q} & \mathbf{0}_{q\times m}  \\
R^{1/2} Y & \mathbf{0}_{m\times q} &\mathbf{0}_{m\times q} &\mathit{I}_{m\times m} 
\end{bmatrix} 
\\
\qquad \qquad \qquad \qquad \qquad  \qquad  \qquad \geq 0, \ \text{for} \ i =1,2,\ldots,l.
\end{gathered}
\label{eq:LMI_conv}
\end{equation}
The problem \eqref{eq:LMI_conv} can be solved by convex optimization software, YALMIP \cite{YALMIP}.
Once the state feedback gain $K$ is obtained, the \Cref{ams:qs} is satisfied, and then the RPI set $\mathbb{S}$ of the error system \eqref{pre:error} can be calculated as follows:
\begin{equation}
\begin{aligned}
    \mathbb{S} = \mathbb{W} &\oplus \text{Conv}\{A^c(p^i)\mathbb{W}, \forall i\in\{1,2,\ldots,l\}\} \\ &\oplus \text{Conv}\{A^c(p^i)A^c(p^j)\mathbb{W}, \forall i,j \in \{1,2,\ldots,l\}\} \\ & \oplus \cdots.
\end{aligned}
\end{equation}

\subsection{Robust MPC strategy}
The nominal control input can be computed using the following MPC problem with the RPI set:
\begin{equation}\label{eq:RMPC}
\begin{gathered}
    \min_{\mathbf{\bar{y}}_{(\cdot)},{\mathbf{\bar{u}}}_{(\cdot)}} 
    \sum_{k=0}^{N-1}( || \mathbf{\bar{y}}_{k|t} ||^{2}_{Q} + || \mathbf{\bar{u}}_{k|t} ||^{2}_{R} )
    + || \mathbf{\bar{y}}_{N|t}||^{2}_{P}, 
\end{gathered}
\end{equation}
\begin{equation}
    \begin{aligned}
        \text{s.t. } 
        & \mathbf{\bar{y}}_{0|t} = \Psi_{\mathbf{x}}(\mathbf{x}_{0|t}), \\
        & \mathbf{\bar{y}}_{k+1|t} = A(p_{k|t})\mathbf{\bar{y}}_{k|t} + B(p_{k|t})\mathbf{\bar{u}}_{k|t},\\  & \qquad \qquad \qquad \qquad k=0,\cdots,N-1,\\
        & C\mathbf{\bar{y}}_{k|t} \in \mathbb{X} \ominus C\mathbb{S}, \ k=0,\cdots,N-1, \\
        & \mathbf{\bar{u}}_{k|t} \in \mathbb{U} \ominus CK\mathbb{S}, \ k=0,\cdots,N-1, \\
        & \mathbf{\bar{y}}_{N|t}  \in \mathbb{Y}_f \ominus \mathbb{S}, \label{eq:constraints}  
        % h  &\leq  E [\bar{d}_{k|t}, \bar{e}_{k|t}, \bar{\psi}_{k|t}]^\top , \label{eq:TMPC_y_C5}
\end{aligned}
\end{equation}
where $N$ is the prediction horizon, $Q$, $R$, and $P$ penalize the state, input, and terminal state, respectively, the subscript $(\cdot)_{k|t}$ represents the value at time $t+k$ predicted at time $t$, and $\mathbb{Y}_f$ is the terminal set.

\begin{definition} [Maximal positively invariant set]
A set $\Omega_{\infty} \subset \mathbb{Y}$ is a maximal positively invariant set (MPI) set of the system $\mathbf{y}_{k+1} = A^c(p_k)\mathbf{y}_k + \mathbf{w}_k$ if $\Omega_{\infty}$ is invariant and all RPI sets are contained. 
\end{definition}

In MPC design, the state feedback gain $K$, obtained from \eqref{eq:LMI_conv} and its corresponding $P$ matrix, can be used to establish recursive feasibility and stability through a terminal set and cost \cite{borrelli2017predictive}. 
The terminal set is obtained by implementing the terminal control input strategy $\mathbf{\bar{u}}_{N|t} = K \mathbf{\bar{y}}_{N|t}$. The set is designed to ensure the satisfaction of the following condition:
\begin{equation}
    \mathbf{y}_{N|t} \in \mathbb{Y}_f \ \Rightarrow \  \mathbf{y}_{N+1|t} \in \mathbb{Y}_f, \  \forall t\in \mathbb{N}^{+}, C\mathbb{Y}_f \subset \mathbb{X}.
    \label{terminal_set}
\end{equation}
The MPI set is often chosen as the terminal set, but in practice, the RPI set can be used if the nominal system \eqref{eq:nominal_system} is stable. 

\subsection{Recursive feasibility and stability analysis}
\begin{assumption} \label{asp:initial}
At the initial time, a feasible solution exists for the nominal PVKO-MPC problem.
\end{assumption} 
\begin{assumption} \label{asp:param}
The model parameter $p_k$ is known over the prediction horizon.
\end{assumption} 
\begin{assumption} \label{asp:cost}
The stage cost and terminal cost are positive definite functions, i.e., they are strictly positive and only equal to zero at the origin.
\end{assumption}

\begin{theorem} \label{thm:feasible}
Assume that \Cref{asp:initial,asp:param} hold. Then, for any time $t$, a feasible solution to the PVKO-MPC problem \eqref{eq:RMPC} always exists.
\end{theorem}

\begin{proof}
Let the initial time be $t$, and let the feasible optimal control sequence and the corresponding state sequence be as follows:
\begin{equation}\label{eq:optcon}
    \begin{gathered}    
        \bar{U}^*_t = [\mathbf{\bar{u}}^*_{0|t},\mathbf{\bar{u}}^*_{1|t},\ldots,\mathbf{\bar{u}}^*_{N-1|t}], \\ 
        \bar{Y}^*_t = [\mathbf{\bar{y}}^*_{0|t},\mathbf{\bar{y}}^*_{1|t},\ldots,\mathbf{\bar{y}}^*_{N|t}].
    \end{gathered}
\end{equation}
At the next time $t+1$, we can obtain the predicted state sequence with the control law $\bar{U}_{t+1} = [\mathbf{\bar{u}}^*_{1|t},\mathbf{\bar{u}}^*_{2|t},\ldots,\linebreak[0]\mathbf{\bar{u}}^*_{N-1|t}, K \mathbf{\bar{y}}^*_{N|t}]$ as $\bar{Y}_{t+1} = [\mathbf{\bar{y}}^*_{1|t},\mathbf{\bar{y}}^*_{2|t},\ldots,\linebreak[0]\mathbf{\bar{y}}^*_{N|t},A^c(p_{N-1|t+1})\mathbf{\bar{y}}^*_{N|t}]$.
Under the \Cref{asp:initial}, the terminal state $\mathbf{\bar{y}}^*_{N|t}$ at time $t$ satisfies the terminal constraints. Then under the condition of the terminal set \eqref{terminal_set}, $A^c(p_{N-1|t+1})\mathbf{\bar{y}}^*_{N|t}$ also satisfies the terminal constraints.
As a result, the MPC problem \eqref{eq:RMPC} is recursively feasible due to the above recursion.
\end{proof}
\begin{theorem}
Suppose that \Cref{asp:initial,asp:param,asp:cost} hold, the system \eqref{eq:nominal_system} is asymptotically stable under the solution to the MPC problem \eqref{eq:RMPC}.
\end{theorem}
\begin{proof}
Let $J_t$ be a Lyapunov function defined as follows:
\begin{equation}
    J_{t} = \sum_{k=0}^{N-1}
    (|| \mathbf{\bar{y}}_{k|t} ||^{2}_{Q} + || \mathbf{\bar{u}}_{k|t} ||^{2}_{R} )
    + || \mathbf{\bar{y}}_{N|t}||^{2}_{P}.
\end{equation}
Let $J^*_{t}$ be the optimal cost at time $t$, which can be computed by \eqref{eq:optcon}, and also let $\hat{J}_{t+1}$ be the cost at time $t+1$, which can be computed by $\bar{U}_{t+1}$ and $\bar{Y}_{t+1}$ as follows:
\begin{align}
    \hat{J}_{t+1} &= \underbrace{\sum_{k=0}^{N-1}
    (|| \mathbf{\bar{y}^*}_{k|t} ||^{2}_{Q} + || \mathbf{\bar{u}^*}_{k|t} ||^{2}_{R} )}_{=J^*_t- || \mathbf{\bar{y}}^*_{N|t}||^{2}_{P}} - \underbrace{(|| \mathbf{\bar{y}^*}_{0|t} ||^{2}_{Q} + || \mathbf{\bar{u}^*}_{0|t} ||^{2}_{R} )}_{\geq 0 \ (\Cref{asp:cost})} \nonumber \\ 
    & \quad  + (|| \mathbf{\bar{y}^*}_{N|t} ||^{2}_{Q} + || K\mathbf{\bar{y}^*}_{N|t} ||^{2}_{R} ) + || \mathbf{\bar{y}}_{N|t+1}||^{2}_{P} \nonumber \\
    & \leq J^*_t - || \mathbf{\bar{y}}^*_{N|t}||^{2}_{P} + || \mathbf{\bar{y}^*}_{N|t} ||^{2}_{Q} + || \mathbf{\bar{y}^*}_{N|t} ||^{2}_{K^\top RK} \nonumber \\ & \quad  + ||\mathbf{\bar{y}}^*_{N|t}||^{2}_{{A^c}^\top PA^c}  \nonumber \\ 
    & \leq J^*_t + \underbrace{|| \mathbf{\bar{y}}^*_{N|t}||^{2}_{ {A^c}^\top PA^c-P+ Q +K^\top RK}}_{\leq 0 \ (\Cref{ams:qs})}  \nonumber \\ 
    & \leq J^*_t,
\end{align}
where $A^c = A^c(p_{N-1|t+1})$.
As $J^*_{t+1}\leq \hat{J}_{t+1}$, we can obtain $J^*_{t+1} \leq J^*_{t}$. Thus, with the proposed controller \eqref{eq:RMPC}, the nominal system \eqref{eq:nominal_system} converges to zero as $t\rightarrow \infty$.
\end{proof}

\section{Simulation results}
The performance of the modeling accuracy and control system was verified through two simulations. 

\begin{figure}[t]
    \centering
    \includegraphics[width=0.95\linewidth]{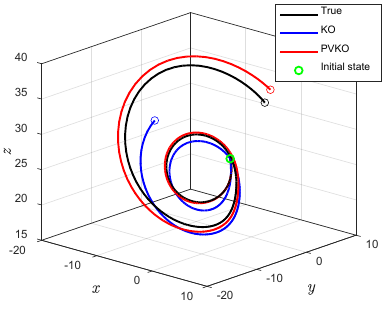}
    \caption{Comparison of predicted trajectories.}
    \label{fig:3d_traj}
\end{figure}

\begin{figure}[t]
    \centering
    \includegraphics[width=0.98\linewidth]{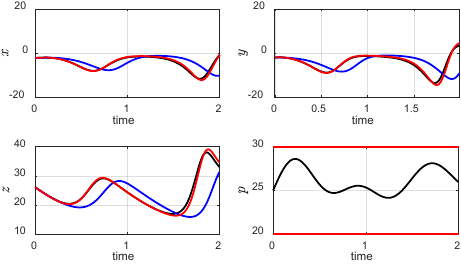}
    \caption{Time trajectories of the states and parameter.}
    \label{fig:model_result}
\end{figure}

\begin{figure}[t]
    \centering
    \includegraphics[width=0.9\linewidth]{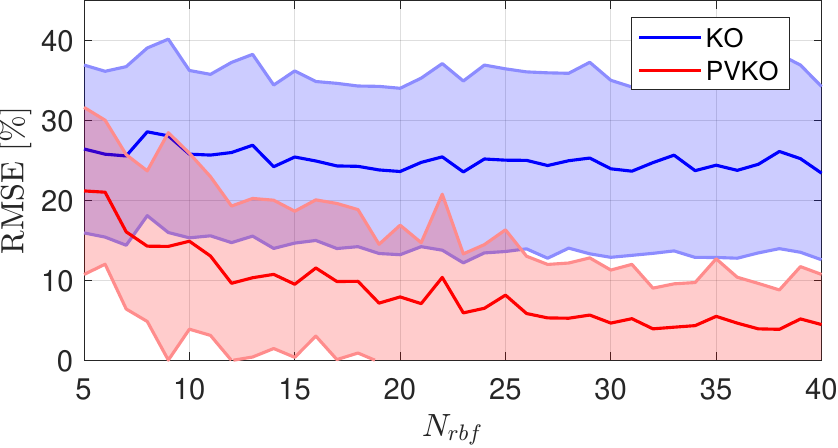}
    \caption{Comparison of performance according to the order of lifting function (The shaded region represents one standard deviation from the mean).}
    \label{fig:model_MC}
\end{figure}

\begin{figure*}[h]
    \centering
    \begin{subfigure}[h]{0.99\linewidth}
        \centering
        \includegraphics[width=\textwidth]{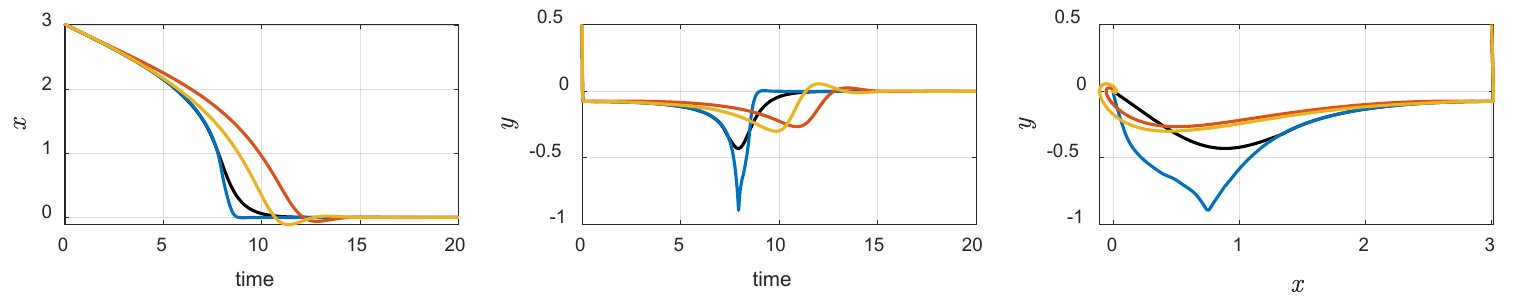}
        \caption{Time trajectories of the states.}
        \label{fig:result2_1}
    \end{subfigure}
    \begin{subfigure}[h]{0.325\linewidth}
        \centering
        \includegraphics[width=\textwidth]{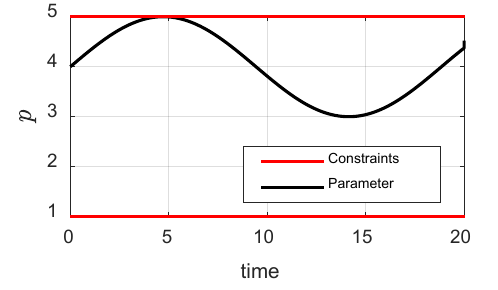}
        \caption{Time trajectory of the parameter.}
        \label{fig:result2_2}
    \end{subfigure}
    \begin{subfigure}[h]{0.325\linewidth}
        \centering
        \includegraphics[width=\textwidth]{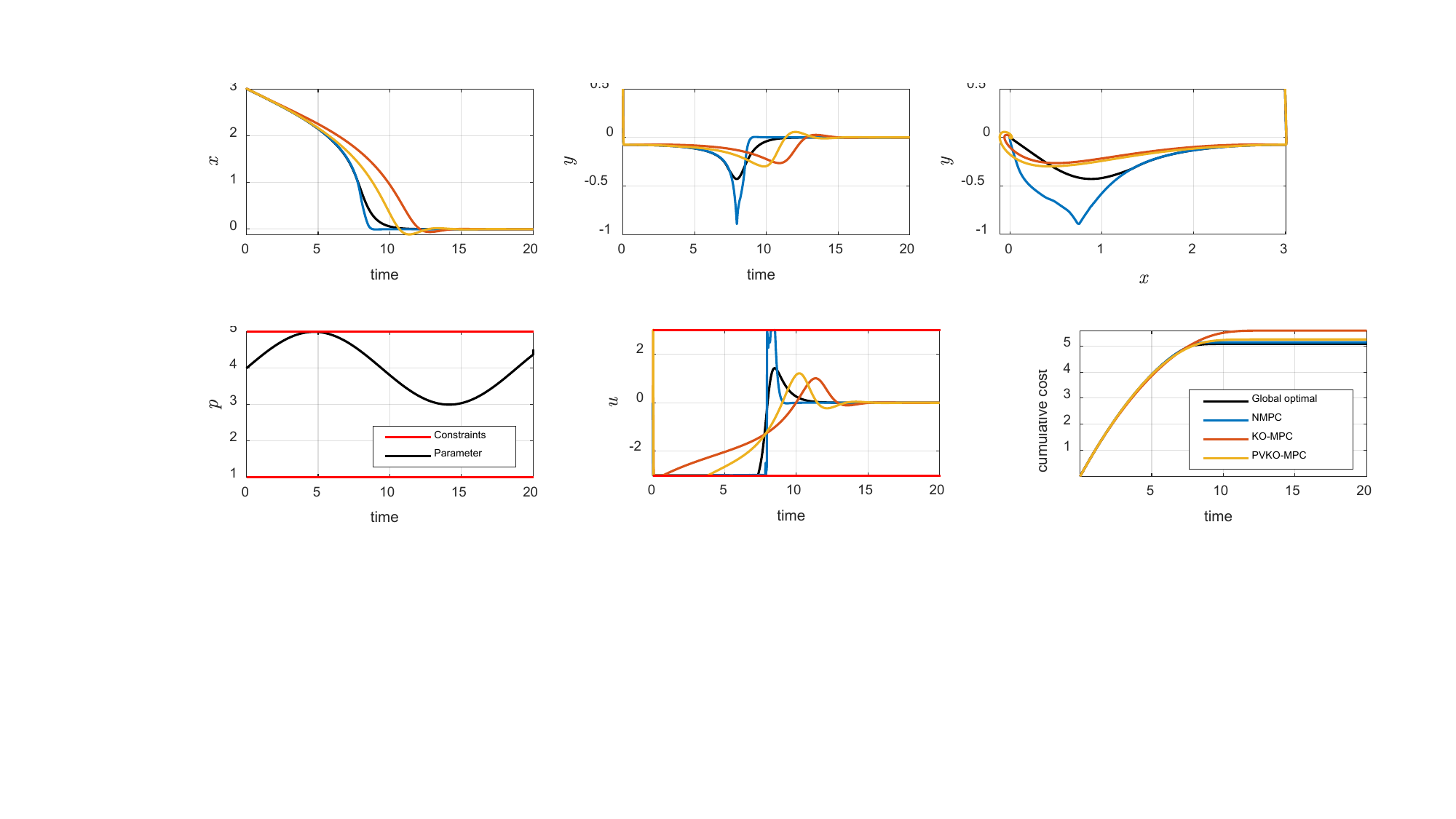}
        \caption{Time trajectories of control inputs.}
        \label{fig:result2_3}
    \end{subfigure}
    \begin{subfigure}[h]{0.325\linewidth}
        \centering
        \includegraphics[width=\textwidth]{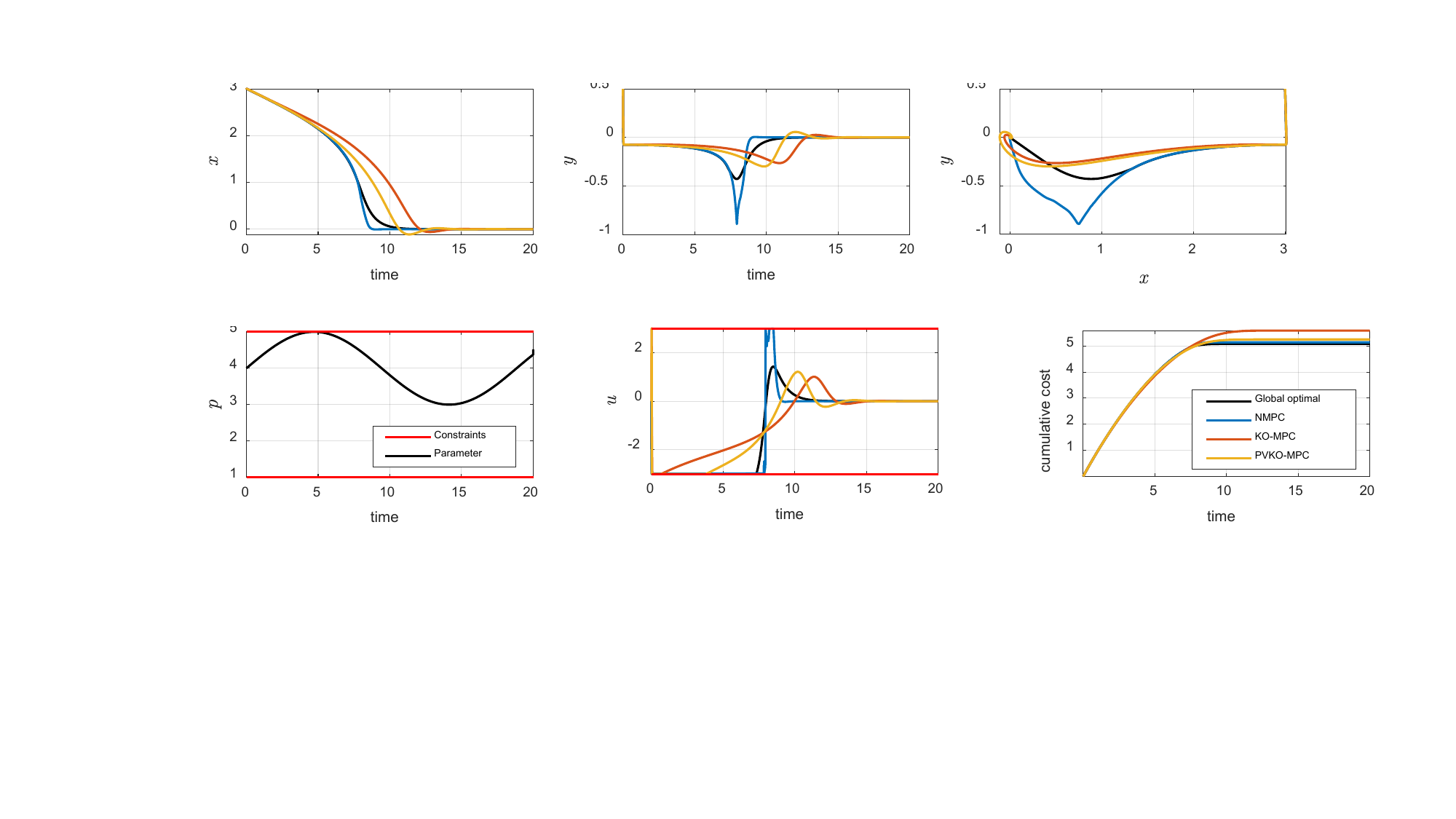}
        \caption{Time trajectories of cumulative cost.}
        \label{fig:result2_4}
    \end{subfigure}
    \caption{Simulation results.}        
    \label{fig:result2}
\end{figure*}

\subsection{Prediction Performance}
The prediction performance of the proposed PVKO approach is evaluated using the Lorenz model, which is defined by the following set of equations:
\begin{equation}
    \begin{aligned}
    \dot{x} &= 10(y -x), \\
    \dot{y} &= px - y - xz, \\
    \dot{z} &= xy - z,
    \end{aligned}
\end{equation}
where $p$ is a time-varying parameter defined as $\textstyle p(t) = 25+\sum_{i=1}^{20} a_i \sin(f_i t)$, where $\textstyle \sum_{i=1}^{20} a_i = 5$ and $a_i>0$. The value of $f_i$ is selected from a uniform distribution from $0$ to $10$. The proposed PVKO approach is applied using 50 thin-plate radial basis functions for the lifting functions, with three working points, $p=20,25,30$. The simulation data with a 50 s duration and a 0.01 s sampling time are used for each working point's KO modeling, while 150 s of simulation data were used for conventional (time-invariant) KO modeling. The prediction is performed for 2 s, and the resulting trajectory and parameter over time are shown in Figs.~\ref{fig:3d_traj} and \ref{fig:model_result}. 

To evaluate the quantitative performance and the effects of the order of lifting function, a Monte-Carlo simulation was conducted. For each order, 500 prediction simulations with a 200-step prediction (2 seconds) were conducted, and the prediction accuracy was computed using the root mean square error (RMSE) as follows:
\begin{equation}\label{eq:rmse}
\text{RMSE}= 100 \frac{
\sqrt{\sum_k ||\mathbf{\hat{x}}_k - \mathbf{x}_k ||_2^2}}
{\sqrt{ \sum_k || \mathbf{x}_k ||_2^2 }},
\end{equation}
where $\mathbf{\hat{x}}_k$ is a predicted state vector.
As shown in Fig.~\ref{fig:model_MC}, the proposed PVKO approach outperforms the time-invariant KO for the parameter-varying Lorenz model simulation.

\subsection{Control Performance}
The performance of the PVKO-MPC is evaluated using the Van der Pol oscillator model with a time-varying model, given by:
\begin{equation}
\label{eq:vdp}
    \begin{aligned}
        \dot{x} &= 2y, \\
        \dot{y} &= -0.8x + p(y - 2x^2y) + u,
    \end{aligned}
\end{equation}
where the control input $u$ and the time-varying parameter $p$ are subject to a random walk model and are constrained to specific value ranges. 
\begin{table}[t]
\centering
% \normalsize
\renewcommand{\arraystretch}{1.3}
\caption {Parameters of controllers} 
\label{tab:MPC_Setting}
\begin{tabular}{c | c | c | c | c | c}
\hline
\textbf{Symbol} & \textbf{Value} & \textbf{Symbol}&  \textbf{Value} & \textbf{Symbol}&  \textbf{Value}\\
\hline
\hline
$N$    &  50 & $T_s$  &  0.01 & $[\underline{p}, \overline{p}]$ & $[1, 5]$\\
$Q$  & $\text{diag}([1, 1])$ & $R$  & 0.1 & $[\underline{u}, \overline{u}]$ & $[-3, 3]$ \\
\hline
$K$  & \multicolumn{5}{c}{
$\begin{gathered}[-0.2036, -0.3152, 0.0117, 5.3363\cdot 10^{-5}, -0.0062, \\ 0.0489,  -0.0147, -4.3624\cdot 10^{-5}, 0.0035] 
\end{gathered}$} \\
\hline
\end{tabular}
\end{table}
The proposed PVKO model is identified by using the polynomial function as a lifting functions, given by $\Psi = [x,y,xy,x^2,y^2,x^2y,xy^2,x^3,y^3]^\top$, resulting in a dimension of 9. 
For the PVKO modeling, five working points with $p=1,2,3,4,5$ are used. A 1000 s simulation data with 0.01 s sampling time were used for each working point's KO modeling, while 5000 s simulation data were used for conventional KO modeling. Linear interpolation is used to construct a complete PVKO model. We compared the performance of the PVKO-MPC algorithm with the KO-MPC \cite{korda2018linear} and nonlinear MPC (NMPC) algorithms. It's worth noting that only the NMPC algorithm utilizes full knowledge of the model. The PVKO-MPC algorithm is compared with the KO-MPC and nonlinear MPC (NMPC) algorithm with full knowledge of the model as follows:
\begin{equation}\label{eq:NMPC}
\begin{gathered}
    \min_{\mathbf{x}_{(\cdot)},{\mathbf{u}}_{(\cdot)}} 
    \sum_{k=0}^{N-1}( || \mathbf{x}_{k|t} ||^{2}_{CQC^\top}  + || \mathbf{u}_{k|t} ||^{2}_{R})  + || \mathbf{x}_{N|t}||^{2}_{CPC^\top}  
\end{gathered}
\end{equation}
\begin{equation}
\begin{aligned}
    \text{s.t. } 
    \mathbf{x}_{k+1|t} &= f_d(\mathbf{x}_{k|t},\mathbf{u}_{k|t},p_{k|t}),
     \ k=0,\cdots,N-1, \\
    \mathbf{x}_{k|t} &\in \mathbb{X}, \ k=0,\cdots,N,  \\
    \mathbf{u}_{k|t} &\in \mathbb{U} , \ k=0,\cdots,N-1, 
\end{aligned}
\end{equation}
where $N$ is the prediction horizon, the weight matrices $Q$, $R$, and $P$ are defined as in \eqref{eq:RMPC}, and matrix $C$ is identified in \eqref{eq:ident_C}. The function $f_d$ is obtained by discretizing the nonlinear function \eqref{eq:nonlinear} using the Euler method with a sampling time of $T_s=0.01 \ \text{s}$.
The controller's parameters are provided in Table~\ref{tab:MPC_Setting}.

To compare the performance of two controllers, simulations were conducted using \eqref{eq:vdp} with an initial state of $[x,y] = [3, 0.5]$ and a time-varying parameter is shown in Fig.~\ref{fig:result2_2}.
The PVKO-MPC problem \eqref{eq:RMPC} was solved using the light-weight sparse quadratic programming solver, qpSWIFT \cite{qpSWIFT}, while the interior point optimizer, IPOPT \cite{IPOPT}, with CasADi software \cite{CasADi} in MATLAB was used for NMPC.

\Cref{fig:result2} shows the result of the three controllers and optimal trajectory obtained by \eqref{eq:NMPC} with $N = \infty$. The cumulative cost is calculated as $\textstyle J_c(k) = \sum_{i=0}^k (|| \mathbf{x}_{i} ||^{2}_{CQC^\top} + || \mathbf{u}_{i} ||^{2}_{R} )$, and the resulting costs are shown in Fig.~\ref{fig:result2_4}. As can be seen, the PVKO-MPC spent less cost than KO-MPC in this simulation and almost similar with NMPC, which uses full knowledge of the model. The average computation time and the cumulative cost are summarized in Table.~\ref{tab:result}. 

\begin{table}[!h]
\centering
% \normalsize
\renewcommand{\arraystretch}{1.3}
\caption {Average computation time, the cumulative cost, and the cost ratio of three controllers (where $J^*_c$ is the cumulative cost of global optimal trajectory)} 
\label{tab:result}
\begin{tabular}{c | c | c |c }
\hline
& \textbf{NMPC}& \textbf{KO-MPC} & \textbf{PVKO-MPC} \\
\hline
\hline
 \textbf{Avg. computation time} & 0.0056  & 0.0032  & 0.0033 \\
\hline
\textbf{Cumulative cost $J_c$} & 5136.7 & 5585.5 &  5246.8 \\
\hline
\textbf{$100 \cdot (J_c - J^*_c)/J^*_c $} & 0.71\% &  9.51\%  &  2.87\% \\
\hline
\end{tabular}
\end{table}

\section{Conclusion}
In this paper, we proposed the data-driven PVKO approach for modeling and controlling parametric uncertain nonlinear systems. Our method involved identifying local Koopman operators at each working point and interpolating them to form a complete PVKO. Furthermore, we designed a PVKO-MPC approach with a robust error-compensation controller, derived through linear matrix inequality, and provided recursive feasibility and stability analysis. 
The efficacy of the proposed approach was demonstrated through simulations, which showed improved accuracy in modeling and performance in controlling for uncertain nonlinear systems.

\bibliographystyle{IEEEtran}
\bibliography{IEEEabrv,main}
\end{document}